# Benchmark 3D eye-tracking dataset for visual saliency prediction on stereoscopic 3D video

**Amin Banitalebi-Dehkordi,[a] Eleni Nasiopoulos,[b] Mahsa T. Pourazad,[c] and Panos Nasiopoulos,[a]**
[a] Electrical Engineering Department and ICICS at the University of British Columbia, Vancouver, BC, Canada
[b] Department of Psychology at the University of British Columbia, Vancouver, BC, Canada
[c] ICICS at the University of British Columbia and TELUS Communications Inc., Canada

**Abstract**. Visual Attention Models (VAMs) predict the location of an image or video regions that are most likely to attract human attention. Although saliency detection is well explored for 2D image and video content, there are only few attempts made to design 3D saliency prediction models. Newly proposed 3D visual attention models have to be validated over large-scale video saliency prediction datasets, which also contain results of eye-tracking information. There are several publicly available eye-tracking datasets for 2D image and video content. In the case of 3D, however, there is still a need for large-scale video saliency datasets for the research community for validating different 3D-VAMs. In this paper, we introduce a large-scale dataset containing eye-tracking data collected from 61 stereoscopic 3D videos (and also 2D versions of those) and 24 subjects participated in a free-viewing test. We evaluate the performance of the existing saliency detection methods over the proposed dataset. In addition, we created an online benchmark for validating the performance of the existing 2D and 3D visual attention models and facilitate addition of new VAMs to the benchmark. Our benchmark currently contains 50 different VAMs.

**Address all correspondence to:** Amin Banitalebi-Dehkordi, University of British Columbia, Vancouver, BC, Canada, V6T 2G9, E-mail: dehkordi@ece.ubc.ca

## 1 Introduction

When watching natural scenes, the Human Visual System (HVS) receives and overwhelming amount of visual data, much beyond its capability to process at once[1]. In order to process these data efficiently, HVS first performs a rapid parallel scan over the scene to locate and prioritize the Regions Of Interests (ROIs). Then, in-depth scene analysis is performed for limited portions of the scene data, according to the priority of different regions. In the literature, the first stage is referred to as "pre-attentive" and the second one is known as the "attentive" stage of vision[2].

The only true way to study the pre-attentive stage of the human vision is to use eye-tracking devices to record the eye movements and extract the eye gazed points (fixations). The information of fixations and saccades collected from these experiments helps us understand the process of ROI detection and prioritization in the HVS. Although eye trackers are used occasionally in some experiments, the use of such devices is not plausible when either an automatic ROI detection is required or setting up the experiment environment and involving subjects is not practical. In this case, computational models of visual attention are employed instead. Visual Attention Models (VAMs) aim to estimate the location of the ROIs by predicting the coordinates of the eye fixation points. VAMs usually generate a Fixation Density Map (FDM) to demonstrate the likelihood for each part of an image or video to attract attention. VAMs are used in a diverse range of applications such as object detection and recognition[3-6], image and video compression[7-9], retargeting and retrieval[10-13], quality evaluation[14-16] and various other fields[17-20].

Visual attention models generally consider a set of saliency attributes and combine these features to an overall saliency map. VAMs are usually classified as either bottom-up or top-down methods according to the saliency attributes they consider. Bottom-up VAMs are the ones that only incorporate the low-level stimulus-driven information of the scene such as brightness contrast, color, motion, texture orientations, or depth. Top-down VAMs, on the other hand, adopt higher-level context-dependent information such as the presence of humans, text, animals, or vehicles for some specific tasks under investigation. In addition to the bottom-up and top-down attention models, there are also some less common integrated models, which combine the bottom-up and top-down saliency attributes[21-22].

Most of the existing VAMs are designed for saliency detection in 2D image and video content. However, natural scenes are seen in a 3D environment. In fact, the perceived depth of a scene is a strong source of information for HVS that can significantly alter the attention towards specific regions in that scene. Therefore, accurate saliency prediction is possible only when depth effects are also taken into account. Moreover, considering that recent advances in 3D technologies make use of consumer 3D systems (such as 3D TVs, cameras, displays, theaters, or games), saliency detection mechanisms incorporated in these systems have to consider the depth perception paradigm of HVS as well as the 2D saliency attributes.

Eye-tracking datasets are usually used as ground-truth to validate the performance of different saliency prediction methods. In a recent study by Engelke et al. it was found that FDMs resulting from independent eye-tracking experiments were very similar[23]. In other words, independently conducted eye-tracking experiments provided similar results for saliency prediction, quality assessment, and image retargeting purposes. This suggests that a benchmark eye-gaze dataset can be robustly used as a reference ground-truth point for various applications. The research community has made publicly available over a dozen 2D image and video datasets, to facilitate testing the performance of automatic VAMs for predicting human fixation points[24-50]. In the case of 3D, however, there are very few stereoscopic image datasets available[51-53]. While there are plenty of eye-tracking datasets available for 2D VAM studies, there is only a couple of stereoscopic video dataset available so far, which contain 8 and 47 stereoscopic sequences[54,138]. The lack of such 3D datasets is an additional obstacle in evaluating and comparing 3D-VAMs and saliency prediction mechanisms for 3D video content.

This paper introduces a large-scale publicly available eye-tracking dataset of stereoscopic 3D videos. Our dataset contains eye-tracking data collected from 61 stereoscopic 3D videos (and 2D versions of those) and 24 subjects. We created an online benchmark system[55] to validate and compare the performance of the existing 2D and 3D visual attention models over our dataset. Our online benchmark currently contains 50 different VAMs and facilitates addition of new VAMs to the benchmark.

The rest of this paper is organized as follows: Section 2 reviews the existing 3D eye-tracking datasets, Section 3 provides an overview of existing 3D-VAMs, Section 4 describes the specifics of our dataset (including 3D video capturing, equipment, subjective experiments, data collection and analysis processes), Section 5 contains the performance evaluation results of the existing VAMs using the proposed dataset, and Section 6 concludes the paper.

## 2 Review of the Existing Eye-Tracking Datasets

Available eye-tracking datasets are categorized as: 2D image, 2D video, 3D image, and 3D video datasets. In this section we introduce the representative datasets for the 3D category. Table 1 provides the specification of the datasets mentioned in this section at a glance. Readers are referred to Ref. 32 for a review on the existing 2D eye-tracking datasets.

### *2.1 3D Image Eye-Tracking Datasets*

There are only a few 3D eye-tracking datasets publically available. The NUS3D dataset collected human eye fixations from 600 stereopairs viewed by 80 subjects[52]. To create this dataset, a

**Table 1.** Description of different 3D eye-tracking datasets

| Dataset | Scenes | Type | Year | Resolution | Subjects | Eye tracker | Sampling freq. (Hz) | Viewing dist. (cm) | Screen diag. (in) | Screen type | Users' age | Ave. length (sec) |
|---|---|---|---|---|---|---|---|---|---|---|---|---|
| 3DGaze database [51] | 18 | 3D Image | 2013 | varying | 35 | SMI RED 500 | 500 | 93 | 26 | LCD | 18-46 | 15 |
| HVEI2013 [53] | 54 | 3D Image | 2013 | 1920×1080 | 15 | Tobii x50 | 50 | 234 | 42 | LCD | 21-60 | 20 |
| NUS3D [52] | 600 | 3D Image | 2012 | 640×480 | 80 | SMI | NA | 80 | NA | LCD | 20-33 | 6 |
| EyeC3D [54] | 8 | 3D Video | 2014 | 1920×1080 | 21 | Smart Eye Pro | 60 | 180 | 46 | LCD | 18-31 | 8-10 |
| IRCCyN 3D Video [138] | 47 | 3D Video | 2014 | 1920×1080 | 40 | SMI RED | 60 | 93 | 26 | LCD | 19-44 | NA |
| Proposed: UBC3Deye | 61 | 3D Video | 2015 | 1920×1080 | 24 | iView X RED | 250 | 183 | 46 | LCD | 20-30 | 8-12 |

Kinect sensor was utilized to capture depth at 640×480 resolution. The depth maps were then used along with a captured color view (left view) to synthesize the right view. The purpose of this study was to validate a visual attention model for 3D stereopairs. The "3DGaze database" is another source for eye tracking information of 3D stereopairs. This dataset contains gaze information of 18 stereoscopic images at various resolutions viewed by 35 users[51]. The stereoscopic images used in this dataset were captured using 3D cameras. Disparity maps were later generated by automatic disparity map generation algorithms. In a study by Khaustova et al.[53], a dataset of stereopairs was introduced that contains eye fixation data corresponding to 54 captured stereoscopic images at full-HD resolution (1920×1080) viewed by 15 subjects. This dataset was then used to investigate the possible changes in visual attention with respect to depth and texture variations.

### *2.2 3D Video Eye-Tracking Datasets*

To the best of our knowledge, to this date, the only publicly available 3D video eye-tracking datasets are the EyeC3D[54] and the IRCCyN[138] datasets, which contain 8 and 47 stereoscopic videos, respectively.

## 3 Overview of the State-of-the-Art 3D VAMs

The overwhelming majority of the existing literature on visual attention modeling is dedicated to 2D VAMs. Due to recent advances in 3D video technologies, there is currently an increasing interest in 3D VAMs. It is worth noting that 2D saliency prediction mechanisms fail to accurately identify the salient regions in 3D images and videos, as they do not incorporate scene depth information[51,58-60]. Depth perception changes the effect of 2D saliency features such as intensity, texture, color, and motion. In addition, several saliency attributes such as depth range and display size and technology solely exist for the case of 3D[61-64]. Fig. 1 demonstrates an example in which state-of-the-art 2D VAMs fail to precisely detect the salient areas in a stereoscopic image (models are applied only to one view). It is observed from Fig. 1 that although some models provide a saliency map close to the human fixation map to some extent, they are still not able to accurately detect the salient regions. For a survey on 2D VAMs readers are referred to Ref. 22, Ref. 65, and Ref. 66.

Similar to the 2D case, the main body of the state-of-the-art 3D VAMs is based on the concept of feature maps[67-68], conspicuity maps[69], center-surround competition[69], and Feature Integration Theory (FIT)[21]. The multi-resolution technique proposed in an early work by Itti and Koch[70] is also frequently used in the design of 3D VAMs, combining different feature values at several resolution levels.

To account for the effect of depth in 3D visual attention modeling of static scenes (stereoscopic images), Maki et al. proposed the direct use of the disparity map as a weighting factor to a 2D saliency map. Their model is based on the assumption that the objects closer to observer draw the viewer's attention more[71]. In similar approaches, Ouerhani et al.[72] and Potapova et al.[73] generated three depth-driven feature maps and combined them with an existing 2D VAM. Park et al.[74] combined the 2D VAM model of Itti[75] with a map of skin detection and visual-comfort-weighted disparity map. Unlike these studies where the disparity/depth was directly used, Niu et al.[59] argued that the closer objects are not necessarily salient to HVS. On the other hand, they proposed to use a measure of depth abruptness for saliency prediction. They adopted two disparity-based conspicuity maps (one from disparity contrast and another one from 3D comfort zone) and combined them to form an overall fixation density map. Lang et al.[52] extracted the depth information directly using a Kinect sensor and developed a probabilistic framework to measure the saliency probability at every depth level and combine the resulted depth saliency with existing 2D VAMs. Iatsun et al.[76] derived the depth information from only monocular depth cues (i.e., only from one view of an image) and incorporated the resulting artificial depth in conjunction with an existing 2D VAM. Wang et al.[51] generated a depth saliency map using a Bayesian approach to be combined with existing 2D VAMs. They verified their approach using the data collected from eye-tracking experiments. Jiang et al.[77] created a depth saliency map (based on foreground and background extraction) and combined it with the 2D VAM of Ref. 78. In another study by Fang et al., DCT coefficients of 8×8 patches of one view and the disparity map were selected to represent features from intensity, color, texture, and depth[58]. These features were combined to produce a saliency map.

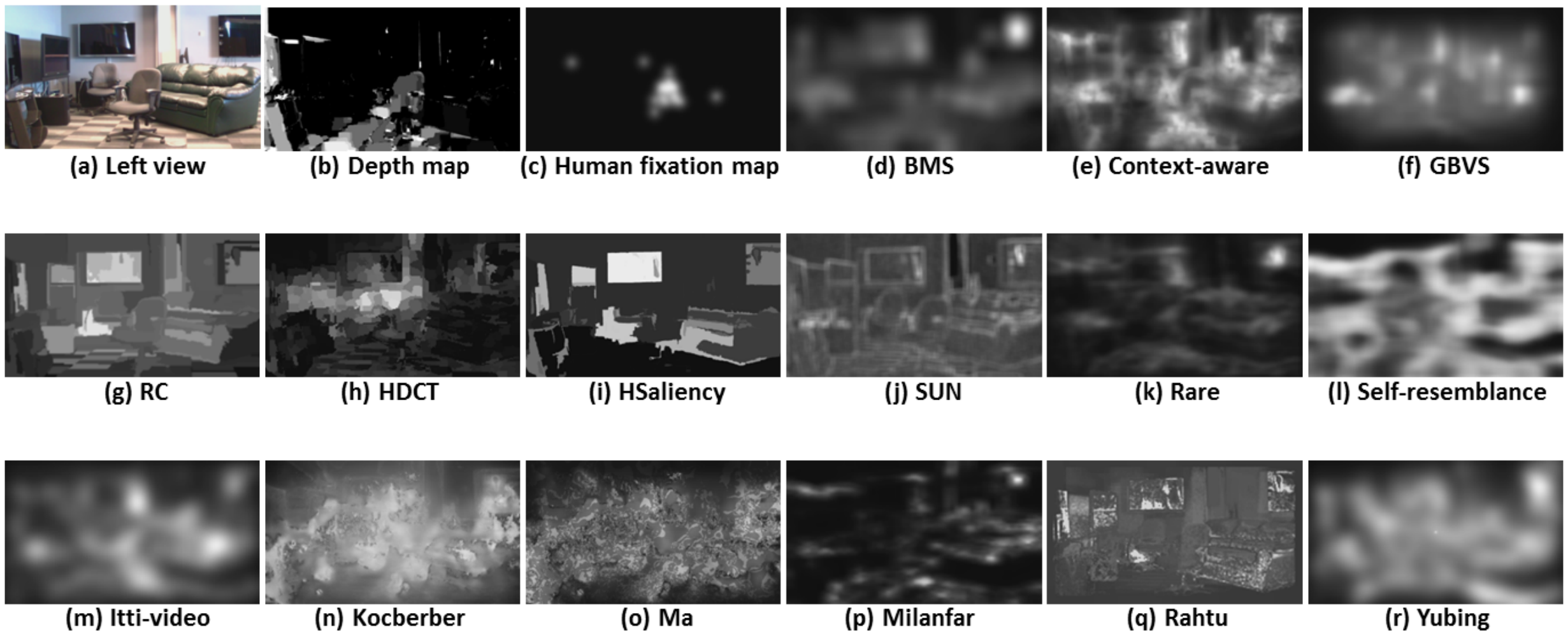

**Fig. 1.** An example showing the inaccuracy of 2D saliency prediction methods when applied to a stereoscopic image pair: (a) The left view of the image, (b) depth map, (c) human fixation map from subjective tests, and generated saliency maps using different methods: (d) BMS[106], (e) context aware[109], (f) GBVS[114], (g) RC[115], (h) HDCT[116], (i) HSaliency[117], (j) SUN[105], (k) Rare[128], (l) self-resemblance static[125], (m) Itti-video[75], (n) Kocberber[121], (o) Ma[123], (p) self-resemblance dynamic[125], (q) Rahtu[127], (r) Yubing[136]. The chair is the most salient object as it stands out in 3D due to its depth.

Compared to 3D VAMs for stereoscopic images, saliency prediction for stereoscopic video has attracted significantly less attention, therefore leaving room for improvement. One reason might be due to the lack of publicly available benchmark stereoscopic video datasets. To account for the impact of depth in 3D video saliency, Chamaret et al.[79] used the disparity information as a weighting factor to refine the ROI selection made by the 2D VAM of Ref. 56. In a similar study by Zhang et al.[80], the depth map is combined with the 2D VAM of Ref. 75 and motion saliency map, to generate a saliency map for 3D video. Coria et al.[81] proposed a reframing framework in which a novel saliency prediction mechanism for 3D video was used. They combined the disparity map (as a conspicuity map) with 2D saliency features to find a bounding box around the most visually important areas in a stereoscopic video. Kim et al.[60] utilized a scene classification approach and incorporated concepts like saliency compactness, depth discontinuities, and visual discomfort for their 3D VAM. They combined the resulting conspicuity maps in a linearly summation or multiplication fashion. In our earlier work[82] we proposed to use a learning-based scheme to developed a random forest model of various low-level saliency features (luminance, color, texture, motion, and depth) as well as the high-level ones (face, person, vehicle, text, animal, and horizon) for saliency prediction on stereoscopic videos. The importance of each feature was reported and the performance of the proposed 3D VAM was validated using a large-scale dataset of stereoscopic videos. Table 2 shows a description of different 3D VAMs at a glance.

## 4 Benchmark 3D Eye-Tracking Dataset for Visual Saliency Prediction on Stereoscopic 3D Video

It is widely acceptable that the existence of a large-scale eye-tracking dataset for stereoscopic 3D videos will accelerate the development of precise 3D VAMs. This paper introduces a publicly available eye-tracking dataset of stereoscopic videos, which aims at facilitating the research community's efforts to compare and validate new 3D saliency prediction algorithms. This section provides information about our eye-tracking dataset and the subjective experiments.

**Table 2.** Existing stereoscopic 3D visual attention models at a glance

| Type | 3D-VAM | Domain | Year | Description | Features | Feature fusion method | Validation dataset |
|---|---|---|---|---|---|---|---|
| 3D images | Maki [71] | Bottom-up | 1996 | Direct use of depth map as a weighting factor in conjunction with an existing saliency detection method | Motion and disparity | Pursuit and saccade modes (AND operator) | Qualitative evaluation |
| | Ouerhani [72] | Bottom-up | 2000 | Use of the model presented in [75] for three depth-based proposed features and combine them | Depth, surface curvature, depth gradient, and color | Weighted average similar to [75] | Qualitative evaluation |
| | Potapova [73] | Bottom-up | 2011 | Three depth-based features are proposed and combined with 2D saliency features | Surface height, orientation, edges, color, and intensity | Average and multiplication | Object labeling accuracy |
| | Park [74] | Bottom-up | 2012 | Combine the 2D VAM model of Itti [75] with a map of skin detection and visual-comfort-weighted disparity map | Intensity, color, orientation, disparity, and faces | Multiplication | Qualitative evaluation |
| | Niu [59] | Bottom-up | 2012 | Two disparity-based conspicuity maps (one from disparity contrast and another one from 3D comfort zone) combined | Disparity | multiplication | Object detection |
| | Lang [52] | Bottom-up | 2012 | Measure saliency probability at every depth Level to combine the resulted depth saliency with existing 2D VAMs | Depth | Average or multiplication | Eye-tracking: 600 scenes and 80 users |
| | Wang [51] | Bottom-up | 2013 | Generate a depth saliency map using a Bayesian approach to be combined with existing 2D VAMs | Disparity | Average | Eye-tracking test: 18 scenes and 35 users |
| | Iatsun [76] | Bottom-up | 2014 | Generate artificial depth from a 2D view and combine it with an existing 2D VAM | Luminance, color, orientation, and depth | Logarithmic multiplication | Eye-tracking test: 27 scenes and 15 users |
| | Jiang [77] | Top-down | 2014 | Create a depth saliency map (based on foreground and background extraction) and combine it with VAM of [78] | The ones in VAM of [78] along with disparity | Weighted average | Subjective quality assessment |
| | Ju [120] | Bottom-up | 2014 | Measure disparity gradient in 8 directions to generate a Anisotropic Center-Surround Difference (ACSD) map from disparity | Disparity | No fusion needed | Object labeling accuracy |
| | Fang [58] | Bottom-up | 2014 | Create feature maps form DCT coefficients of 8×8 patches | Color, luminance, texture, and depth | Weighted sum according to compactness | Eye-tracking test: 18 scenes and 35 users |
| | Fan [112] | Bottom-up | 2014 | Depth, brightness, color and spatial compactness are measured, and combined to generate the region-level saliency map | Luminance, color, and disparity | Summation | Object labeling accuracy |
| 3D videos | Chamaret [79] | Top-down | 2010 | ROIs are detected using the 2DVAM of [56] and refined by the disparity map | Luminance, color, motion, and disparity | Multiplication | ROI detection accuracy |
| | Zhang [80] | Bottom-up | 2010 | Combine depth map with an existing 2D VAM of [75] and motion saliency map | Features of [75] along with motion and disparity | Weighted average | Qualitative evaluation |
| | Coria [81] | Top-down | 2012 | Linear combination of the disparity map (as a conspicuity map) with 2D saliency features | Texture, disparity, motion, color, intensity | Average | Reframing performance |
| | Kim [60] | Bottom-up | 2014 | Combine "saliency strength" for different features taking into account size, compactness, and visual discomfort | Luminance, color, disparity, and motion | Weighted sum | Eye-tracking: 5 scene types and 20 users |
| | LBVS-3D [82] | Bottom-up | 2015 | Learn the important saliency features by training a Random Forest model for saliency prediction | Luminance, color, texture, motion, depth, face, person, vehicle, text, animal, horizon | Random forest learning method | Eye-tracking: 61 scene types and 24 users |

## *4.1 Capturing the Stereoscopic 3D Videos*

Sixty-one (61) indoor and outdoor scenes are selected to produce our 3D video dataset for the eye-tracking experiment. A stereoscopic 3D camera (JVC Everio) was used to capture the sequences. The distance between the two lenses of the camera is 6.5 cm, which is the same as the average inter-ocular distance for humans. The camera was fixed on a tripod at the time of capturing so there were no camera movements. No zooming was used for the camera lenses. Special attention was put during the capturing on avoiding possible window violations. Note that window violation is a common artifact for 3D video acquisition and occurs when parts of an object fall outside the capturing frame. As a consequence, part of the object is perceived inside and screen and other parts appear outside the 3D screen. The two contradictory depth cues result in 3D visual discomfort and reduce the 3D Quality of Experience (QoE).

Each view is captured in full HD (high definition) resolution of 1920×1080 at the frame rate of 30 frames per second (fps). Moreover, the length of each video is approximately 10 seconds. The video sequences contain a wide range of intensity, motion, depth, and texture density. Scenes are selected in a way that the captured dataset covers almost all different possible combinations of these parameters. In other words, there are multiple video sequences captured for each combination of four parameters, intensity, motion, depth, and texture density, with two levels of each (low and high), resulting in 16 different scenarios. As a result, our dataset is not biased towards a specific scenario. These capturing parameters are particularly chosen because they are the key low-level attributes for visual attention modeling[75]. Fig. 2 provides a statistical overview of the video sequences. In this figure, in order to measure the motion and texture density, we utilize the temporal and spatial information (TI and SI, respectively) metrics. We follow the recommendation of ITU[83] for the calculation of SI and TI measures. Applying the SI and TI metrics on the depth maps (instead of the videos) provides us with the Depth Spatial Indicator (DSI) and Depth Temporal Indicator (DTI), respectively[139]. DSI and DTI assess the depth variations of the video content in the spatial and temporal domain.

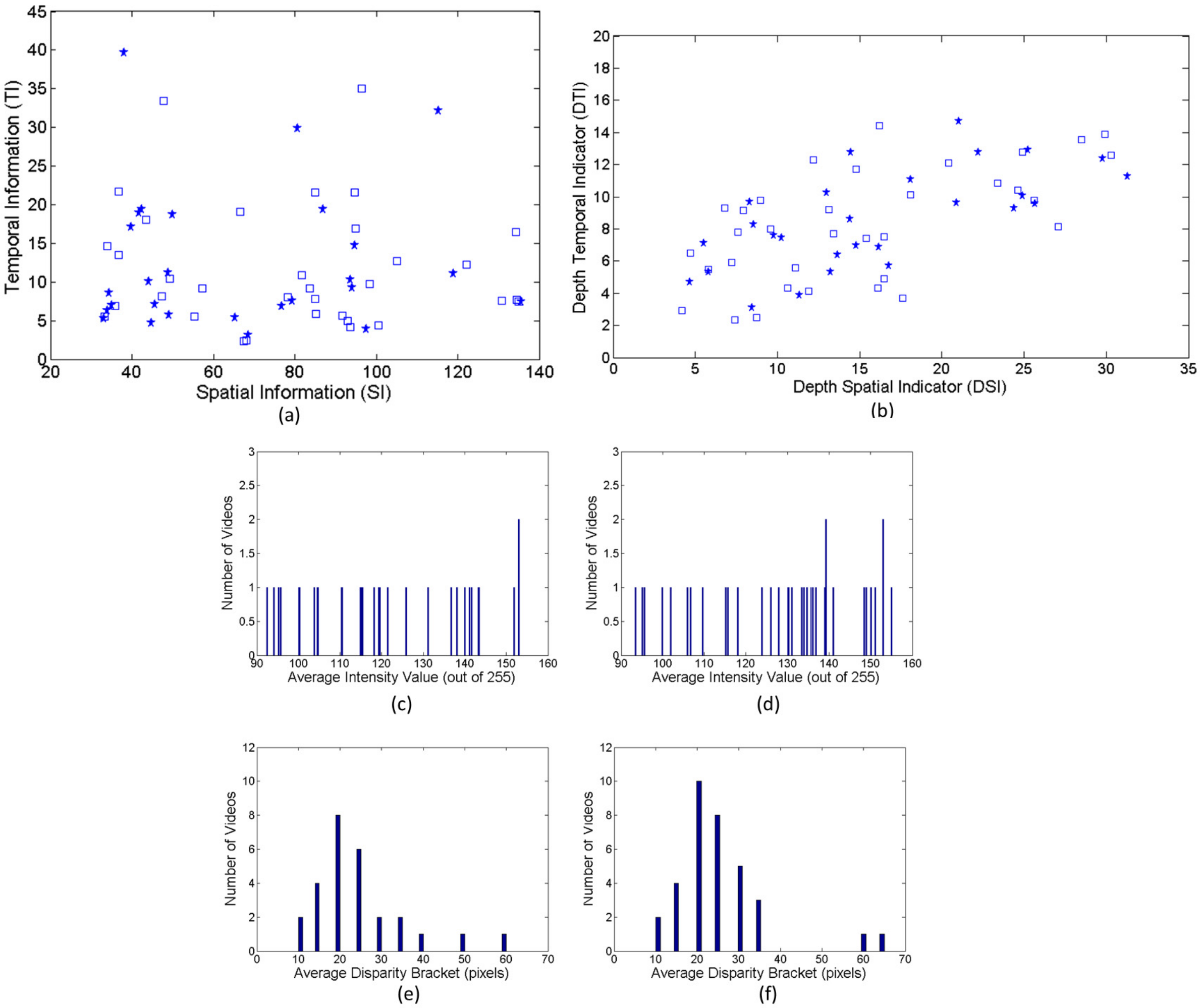

**Fig. 2.** Statistics of the stereoscopic video dataset: Temporal and spatial information measures (a), Depth temporal and spatial indices (b): stars correspond to the training videos and squares refer to validation videos, histograms of the average intensity values for the train set (c) and validation set (d), and histograms of the average disparity bracket for the train set (e) and validation set (f)

In addition, we provide the histograms of the average intensity values for the video dataset as well as the average disparity bracket of the scenes. Disparity bracket is defined as the range of the disparity (in pixels) for each stereopair (i.e., horizontal coordinate of the closest point subtracted by the horizontal coordinate of the farthest point). Note that since some of the VAMs in the literature utilize machine learning techniques for saliency prediction, we split the dataset to the training and validation sets so that the training part can be used for training different VAMs and validation is used for validation of the models. We chose 24 videos for training (~ 40 %) and 37 sequences for validation (~ 60 %). We tried to distribute the videos in the two sets fairly, according to their TI, SI, DSI, and DTI values, intensity histograms, and depth bracket histograms. Statistical properties are provided for both the training and validation sets in Fig. 2. It is observed from this figure that the training and validation sets roughly demonstrate similar statistical properties and, therefore, an unbiased generalization is provided for learning methods.

Moreover, in order to have a balance with respect to the high-level saliency attributes, videos are captured such that humans appear in roughly half (~52 %) of the scenes while in the other half there is no human appearance. Similarly, it is ensured that approximately half of the scenes contain vehicles (~40 %). Note that the low-level capturing parameters are tried to appear evenly for both cases (human vs. non-human). In other words, the video database contains roughly the same combinations of the low-level parameters with and without the appearance of humans (or vehicles). The 61 videos in our dataset are selected from an initial set of over 90 captured videos. A snapshot of the right view of some of the videos in our database is demonstrated in Fig. 3.

Since the videos are captured in two left and right views, disparity estimation algorithms should be incorporated to extract disparity. We provide the disparity maps generated using the Depth Estimation Reference Software DERS[84] of MPEG. Any other method can be used when a disparity map is needed.

*4.2 Post-processing the Captured Videos*

Using a 3D camera with parallel lenses results in negative parallax, which corresponds to objects popping out of the 3D display screen. Experimental studies have shown that viewers distinctly prefer to perceive the objects of the interest on the screen and other objects inside/outside of the screen[85]. At the time of capturing, we used the manual mode of disparity adjustment so that the camera doesn't automatically change the disparity.

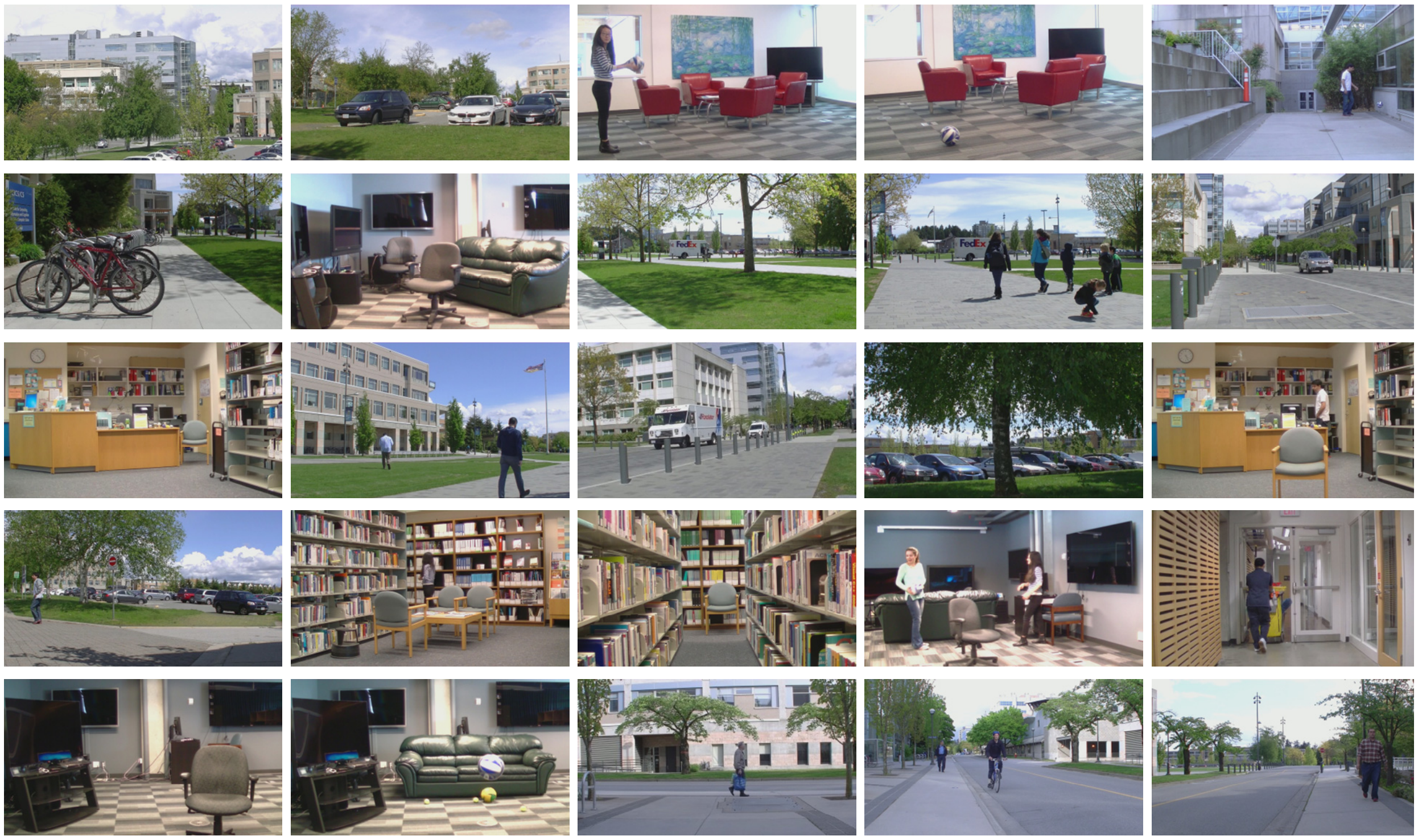

**Fig. 3.** Snapshots of the videos in our database

Therefore, it is required to bring the objects to the 3D comfort zone to provide the viewers with a high 3D quality of experience. To this end, we perform disparity correction to the captured sequences. Disparity correction is achieved by cropping the left side of the left view and the right side of the right view. The amount of cropping is selected based on the disparity of the object of interest in a way that the disparity corrected stereopair places the object of interest on the screen. For each video sequence the object of the interest is identified through a subjective user study. More details regarding this disparity correction methodology can be found in Ref. 86.

### *4.3 Eye-Tracking Experiments*

In order to record the eye movements of the viewers we use a SensoMotoric Instrument (SMI) iView X RED system[87]. This eye-tracker uses infrared illumination to track the eye movements. Images of the eye are analyzed in real-time by detecting the pupil, calculating the center, and eliminating artifacts. To this end, one or several corneal reflexes are tracked in order to compensate for changes in position of the head relative to the camera. The sampling rate of the eye-tracker is 250 Hz, i.e., it has the capability of tracing the eye movements 250 times in each second. Moreover, the accuracy is up to $0.4 \pm 0.03^{\circ}$ of the visual angle (tracking resolution: 0.03° and Gaze position accuracy: 0.4°). The accuracy is achieved with a gaze angle of up to 40° (+/- 20°) horizontally and 60° (+20°/-40°) vertically[87]. A 46" Hyundai (S465) 3D TV was used for displaying the test material. This TV set utilizes passive glasses in the 3D mode. The display LCD size is 57.25 cm vertically and 101.84 cm horizontally. The distance between the TV and the eye-tracker is around 123 cm. The distance between the TV and observers is about 183 cm (3.2 times the display's height). The Fields Of View (FOV) are approximately 32 and 18 degrees, which align with the recommendations in ITU-R BT.2022[140].

Note that the eye-tracking system is claimed by its manufacturers to work robustly in the presence of most contact lenses or eye glasses[87]. However, before the actual experiments, we also verified in our own pre-tests that having eye glasses has no influence on the recorded gaze points. None of the subjects was wearing contact lenses.

The eye-tracker used in our experiments records the movements of both eyes. To create a 3D FDM for each frame, we keep the right eye fixation point as is, but transfer the left eye fixation point to the right eye using the left-to-right disparity map generated by the MPEG Depth Estimation Reference Software (DERS[84]). These two fixation points are later used for creating the FDMs using a Gaussian kernel (sub-section 4.4). Note that a similar approach was used in the creation of the state-of-the-art 3D eye-tracking datasets[51,54,138]. This process results in a fixation map associated to the right eye. The reason of choosing the right eye is that for our 2D experiments we use the right views and also approximately 70 % of humans are right eye dominant[141] (so chances are likely that right eye fixations are better representatives). This process was shown through experiments performed in our lab to offer accurate tracking in 3D content, which was confirmed by comparing the generated fixation maps with the tested 3D content.

The resolution of the display is full HD (1920×1080) in the 2D mode and, thus, there is no need for up/down sampling the video data. Note that the video sequences were displayed in interlaced format so that the eye-tracker grabs the displayed frames through an HDMI cable. Synchronization between the video player and the eye-tracker was ensured, as the eye-tracker utilizes a frame- grabber device to capture the frames shown to the observers. The frame-grabber is directly connected to the TV using an HDMI cable, and takes a copy of the signal presented to the viewers. Peak luminance of the LCD screen is 120 $cd/m^2$. Color temperature was fixed at 6500K, which is recommended by MPEG for subjective evaluation of the proposals submitted in response to the 3D Video Coding Call for Proposals[88]. Subjective experiments were performed in a room that is specifically designed for visual experimental studies. The wall behind the display was illuminated using a uniform light source, with light level of 5% of the screen peak luminance.

In order for the eye-tracker to accurately follow the eye movements, it was placed between the viewers and the 3D display. The distance and height of the eye-tracker is based on the recommendations of the SMI internal software. Subjects were instructed to try to keep their head movements at minimum. The eye movement of the viewers is continuously monitored to ensure that the eye-tracker is constantly recording valid gazed points. The eye-tracker is capable of tracking the viewers' heads at a fairly wide range of 40 × 20 cm at 70 cm distance. Fig. 4 sketches the test setup and placements of the devices. Fig. 5 shows the real test environment.

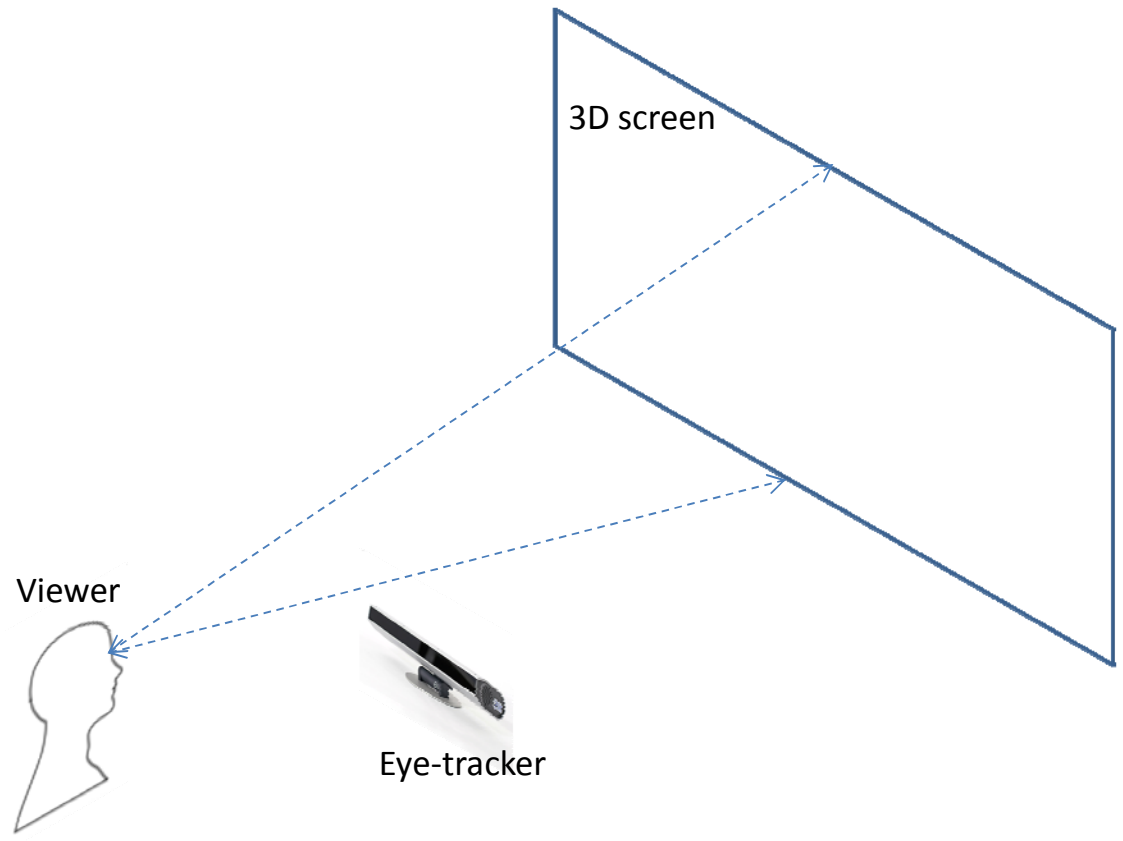

**Fig. 4.** Eye-tracking experiment setup

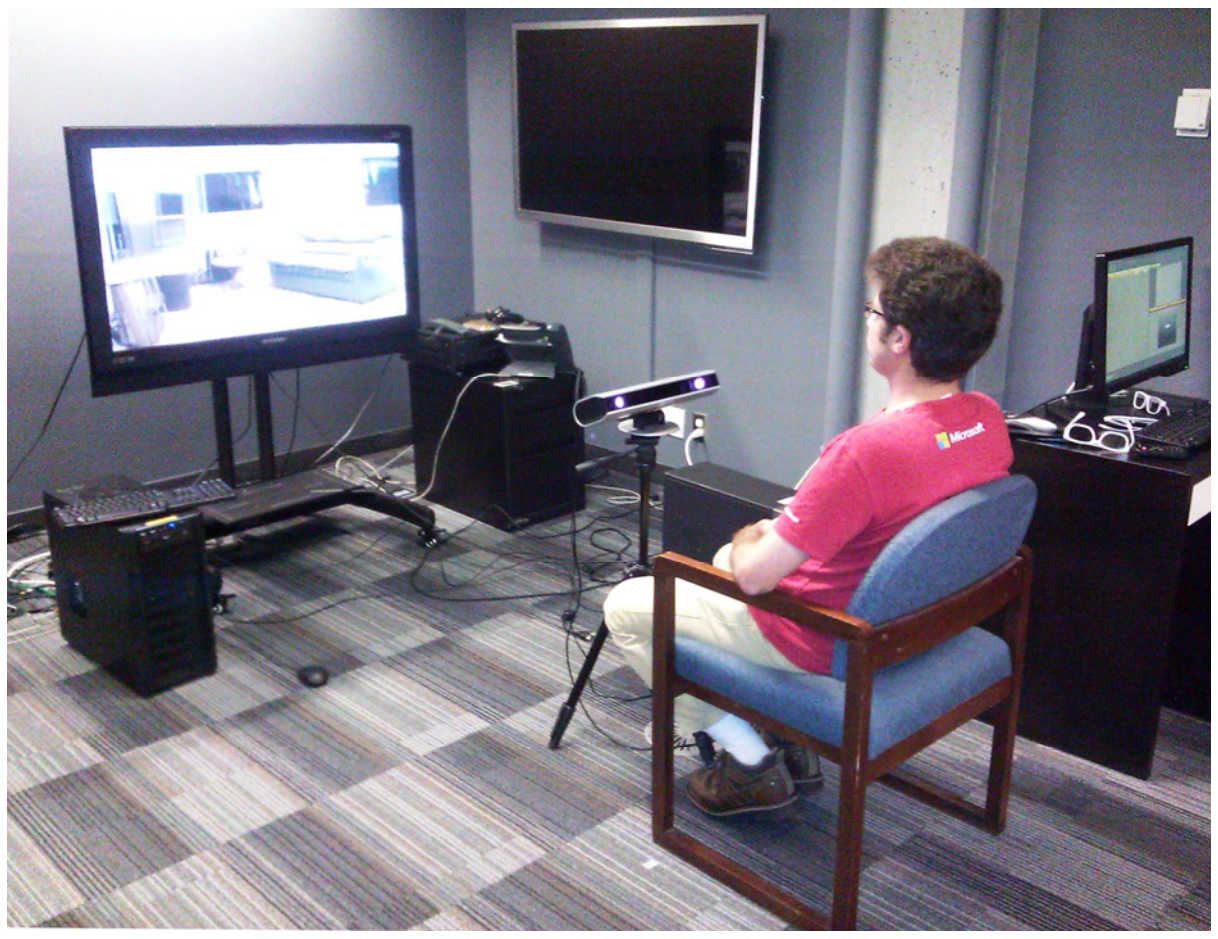

**Fig. 5.** Test environment

Twenty-four participants attended our test sessions (12 male and 12 female). Prior to the actual eye-tracking experiments, all the subjects were screened by pre-tests to assess their visual acuity using the Snellen chart, color blindness using the Ishihara graphs, and stereovision using the Randot test. Users who didn't pass any of the three pre-tests were not entered to the eye-tracking experiment.

Users were selected among university students who were naïve to the purpose of the test and it was ensured that they do not have prior knowledge about the research topic.

Subjects participated in the eye-tracking experiment one at a time. The test involved a free-viewing task, in which viewers freely watched the sequences and their eye movements were recorded. In order to study the statistical differences between the 2D and 3D fixations and saccades, for each participant we performed a test with the 3D sequences and another one with only one of the views (right view). The order of the videos was randomly changed for each subject and each session to avoid memorization of the sequences. Note that having a large number of videos (61) makes it very difficult to memorize the content of different scenes. Also, there was 10 minutes resting (and memory refreshing) period for each subject between the two sessions. To ensure that the subjects were not biased towards the order of the presentation material, the order of the 2D and 3D videos was switched for every subject so that counter balance was achieved.

Prior to each test session, a calibration was performed for each user to ensure the accurate recording of the gaze points. The calibration was performed a few times at the beginning and throughout the experiment so that the eye-tracker doesn't lose track of the eye movements. For calibration, the internal SMI software was used to show a circular dot on the screen and viewer was asked to follow its movement. The circular dot moved in random fashion to the corners of the screen to ensure a reliable calibration for each subject.

### *4.4 Data Collection and Fixation Map Generation*

After performing the eye-tracking experiments, the gaze points are collected and are ready to be processed. In the literature, the fixation points are usually converted to a fixation density map (or sometimes heat map) to provide a 2D representation of most salient points. This map basically shows the likelihood of an object to draw attention. Note that the eye-tracking system used in our experiments already discriminates the fixations and saccades using its internal software. We use only the fixation points in the generation of the FDMs.

Due to the possible inaccuracies of the eye-trackers and to account for the drop in vision sharpness as the distance increases from the fovea center point, the fixation points are usually filtered using a Gaussian kernel to create the fixation density maps[24,51,52,58,89,90]. The Gaussian kernel mimics the photoreceptor distribution of the fovea. Due to photoreceptor distribution in the human eyes, objects at the center of focus are projected to the center of fovea and therefore appear to be the sharpest. As the eccentricity from the fovea increases, vision sharpness drops rapidly[91-94]. The radius size L of the mask is selected based on the 3D display size and the distance of the viewer from the display as follows (Fig. 6):

$$L = Z_{observer}\ \tan(\alpha)\ [cm] = \frac{Z_{observer}\ \tan(\alpha) R_H}{H}\ [pixel] \qquad (1)$$

where H and $R_H$ are the vertical height and resolution of the display and α is the half of the visual angle, which is usually chosen to be one degree of the visual angle [51,58]. In our setup, the distance of the viewer from the display, $Z_{observer}$, is set at 183 cm and the display height, H, is 57.25 cm, which corresponds to a radius of 60 pixels for the resulting Gaussian kernel (The Fields Of View (FOV) are approximately 32 and 18 degrees, which align with the recommendations in ITU-R BT.2022[140]).

After applying the Gaussian kernel to the gazed points, a fixation density map is generated for each frame of the videos. A similar process is repeated for both 2D and 3D videos to generate the fixation maps for each user. User fixation maps are then averaged to create a fixation map video for each sequence. Fig. 7 shows the average of the fixation maps for the 2D and 3D sequences and for all users. It is observed from Fig. 7 that the fixations are distributed all around the frames and therefore there is no significant center-bias observed. This shows that the scenes are captured in a way that viewers are not directed only towards the center of the screen. It also makes the comparison of different VAMs fairer.

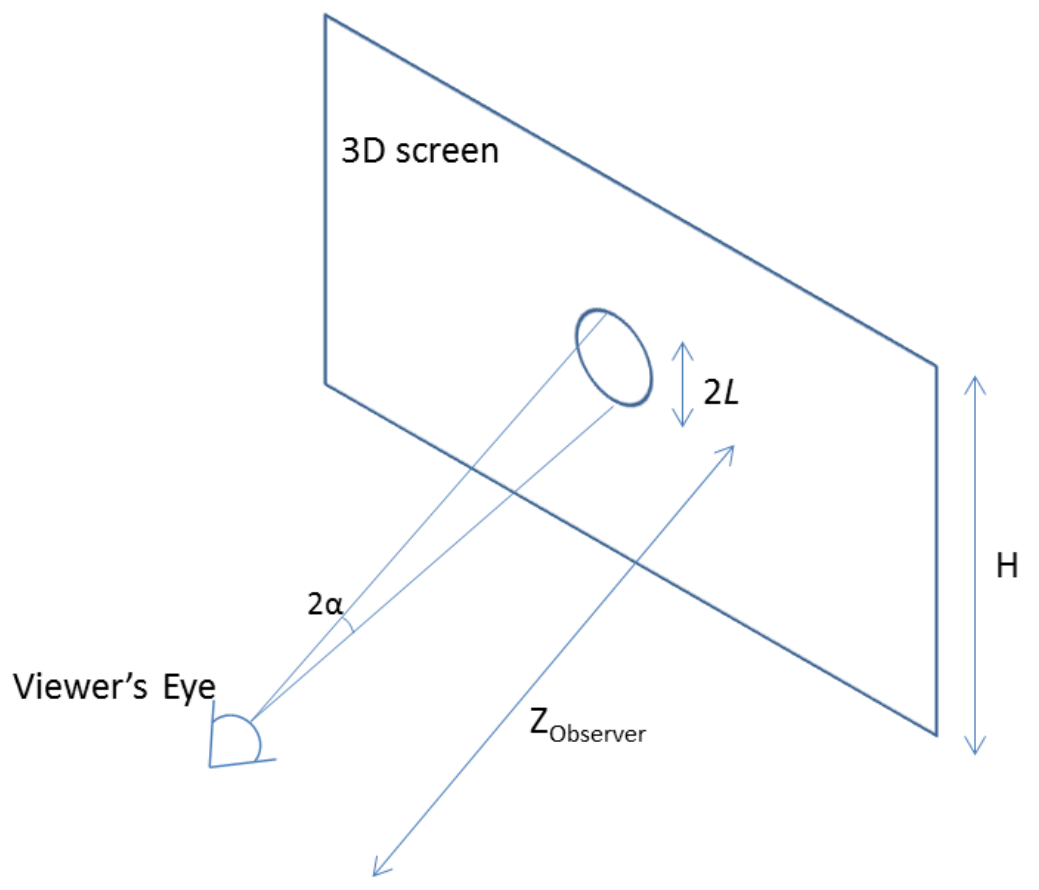

**Fig. 6.** Fovea masking scheme

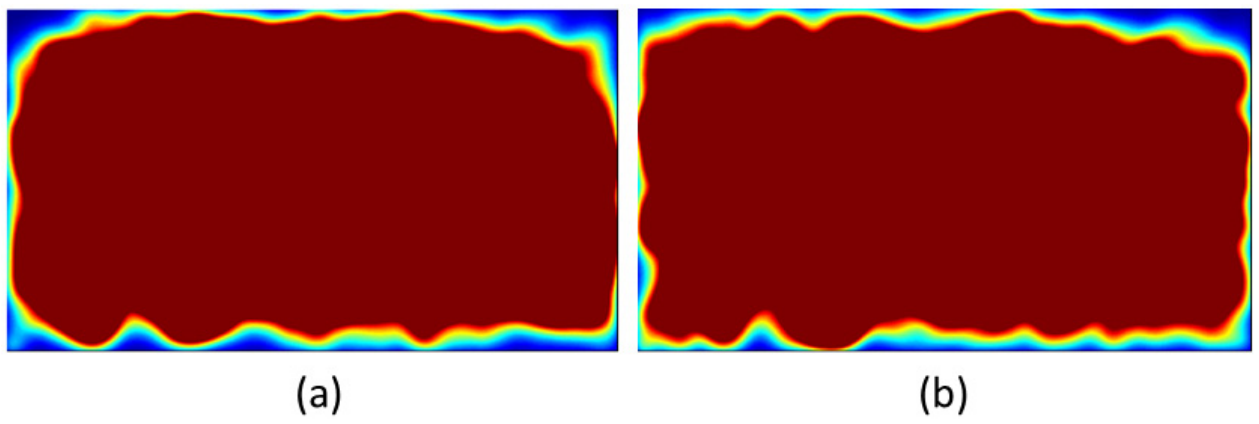

**Fig. 7.** Average fixation maps: 2D videos (a) and 3D videos (b)

The created fixation maps are used as ground truth saliency maps for validating the performances of visual attention models. Our eye-tracking dataset of stereoscopic videos is made publicly available and can be accessed at: http://ece.ubc.ca/~dehkordi/saliency.html. Moreover, we created an online benchmark dataset and evaluated the performance of the existing 2D and 3D VAMs on our 3D eye-tracking dataset. In our online benchmark, we provide performance comparisons of many saliency detection algorithms and facilitate the addition of new models. We provide ground truth fixation maps for the training part of the dataset and ask interested contributors to send us the implementation of their VAM or their generated saliency maps so we can conduct performance evaluations using their input and our validation part of the dataset. The results will be added to the benchmark.

## 5 Performance Evaluations

This section provides details of the currently implemented VAMs in our benchmark and their performance evaluation. Specifically, descriptions of the performance metrics, baseline saliency maps, and comparison of existing VAMs over our eye-fixation dataset of 3D videos are presented.

### *5.1 Metrics of Performance*

The Area Under the Receiver Operating Characteristics (ROC) Curve (commonly referred to as AUC) is a basic measure of performance in our evaluations[95]. In addition to AUC, several other metrics are incorporated for evaluating different VAMs. In particular, the shuffled AUC[96],

Normalized Scanpath Saliency (NSS)[66], Pearson correlation ratio (PCC)[22], Earth Mover's Distance (EMD)[97,98], Kullback–Leibler Divergence (KLD)[99], and a Similarity score proposed by Judd et al.[31] were used in our evaluations. The values of the above metrics are calculated for each frame of our validation video dataset and then averaged over all the frames to provide one metric score for each video. Each of these metrics is explained in this sub-section.

1) AUC: AUC has been frequently used for evaluating the performance of saliency prediction models[95,100-103]. AUC is computed between a saliency map and a ground truth fixation map and quantifies the ability of a saliency map in classifying the fixated and non-fixated locations. AUC achieves values between 0.5 and 1. A higher value of AUC denotes a more accurate saliency map. A value of unity indicates 100% accuracy in saliency prediction and a value of 0.5 indicates that the saliency map predicts the saliency points no better than by chance.

2) Shuffled AUC: AUC is often being criticized for being affected by the center-bias phenomenon[104,105]. To address this disadvantage, the definition of the "negative set" in classic AUC is modified to account for the center-bias problem. The new definition of AUC is known as Shuffled AUC (sAUC)[104,105]. sAUC achieves higher reliability compared to classic AUC and reduces the effect of the center bias. We use a publicly available implementation of sAUC provided by Zhang and Sclaroff[106].

3) NSS: Normalized Scanpath Saliency (NSS) is a measure of how well a saliency map can predict the fixations along the scanpath of the fixations. It is measured as the average of saliency map values along the viewer's scanpath[66]. A saliency value of greater than unity indicates that the saliency map detects saliency much better along the scanpath compared to the other locations. A negative NSS value, however, indicates that the saliency predictions are no better than identifying the salient locations just by chance.

4) PCC: The Pearson Correlation Coefficient (PCC or PLCC) measures the correlation between a saliency map and a fixation map by treating them as two random variables[22]. It is a linear measure of correlation that ranges from -1 to 1, with values close to -1 and 1 indicating a perfect linear correlation (thus a more accurate saliency map) and values near 0 indicating no correlation between a saliency map and a ground truth fixation map (thus very low accuracy).

5) EMD: It is argued that the AUC metric does not take into account the distance between the points in a saliency map and the corresponding points in the fixation map. However, the distance between these points can also be a measure of how close a saliency map is to a fixation map[107]. The Earth Mover's Distance (EMD)[97] is often used for taking into account the distance between two probability distribution functions. In the context of visual attention modeling, EMD measures the cost of converting a saliency map to its corresponding human fixation map. Here, the cost for each window means the difference in the saliency probabilities weighted by the distance between the windows. EMD is a distance measure, so an EMD value of 0 means 100% accuracy of saliency detection, while higher EMD values correspond to lower accuracies. A fast implementation of EMD prepared by Pele and Werman[98] is used in our saliency benchmark.

6) KLD: The Kullback-Leibler Divergence (KLD) (a.k.a information divergence) evaluates the saliency prediction accuracy in an information theoretic context. It models the saliency and fixation maps as two probability distributions and measures how much information is lost when one is used to approximate the other[99]. Similar to EMD, KLD is also a distance measure and achieves non-negative values. Lower KLD values indicate better saliency prediction accuracies.

7) SIM_measure: This Similarity Metric (SIM) was originally used by Judd et al.[31] to evaluate the similarity between a saliency map and its corresponding fixation map. SIM measures the summation of the minimum values of two probability distribution functions (PDFs) evaluated at different points of the distributions. The distributions are scaled so that they sum up to unity. Therefore, SIM takes values in the interval of 0 and 1, 1 indicating that a saliency map perfectly matches the corresponding fixation map, while 0 showing no similarity between the two.

It is worth noting that none of these objective metrics would perform as accurate as human observers. However, except sAUC, which was an improvement over AUC, each of the other metrics was proposed as an attempt to improve the measurement accuracy of VAMs. It was shown in [96] that most saliency prediction performance metrics are similar and that using sAUC, KLD, and one other metric among NSS, PLCC, SROCC, and SIM provides a fair comparison. Although we provide the performance evaluation results of the entire set of metrics for various 3D VAMs, we only use sAUC, KLD, and NSS for ranking their performance.

### *5.2 Baselines*

When comparing the performance of different visual attention models over our stereoscopic video dataset, we use four different baselines as reference models to be compared against each VAM:

1) Chance: a map of random values between 0 and 1 is created to indicate the chance map.

2) Center: This model is particularly chosen to measure how much center-bias exists in the eye-tracking data. A Gaussian kernel with the size of the frame (i.e., 1920×1080 for our dataset) and the optimum standard deviation (STD) is chosen to represent the center map (the kernel is normalized to be 1 at the center and lower values around the center). The optimal value of the standard deviation is chosen by measuring the AUC between the resulting center map and the fixation maps of the training set for many different STD values. The value of STD is exhaustively swept between 0.1 to 1000 (using MATLAB's *fspecial* function); the optimal value of STD=300 resulted in AUC of 0.6064 over the validation set.

3) One human's opinion: since people look at different parts of videos, the fixation distributions are different for different viewers. As a result, the fixation map of one viewer can only partially reflect the true average human fixation map. As a baseline saliency model, we measure how well one human observer can predict the fixation map of the other observers. In other words, the fixation map of each participant is used as a saliency map and its performance in predicting the fixation maps of other participants is evaluated. This process is repeated for all the subjects and the average performances are used for the baseline.

4) Infinite human observers: ideally, finding the average of the fixation maps of a large number observers results in the most reliable and accurate representation of the human fixation map. In practice, there would be only a limited number of participants in a test. Therefore, we find an estimate of the performance for infinite number of humans and use that as an upper bound for the performance of visual attention models. To this end, we find the performance of $i$ humans in predicting the fixations of $N$-$i$ humans. Note that $N$ is the total number of participants (which is 24) and $i$ can vary from 1 to 12 (12 observers predicting the fixations of the other 12 observers). For each value of $i$, we repeat the process for all possible combinations of $i$ participants ($i$ out of 24) to ensure more robustness (less bias towards particular subjects). Fig. 8 shows the performance of the observers in predicting fixations of other observers. The AUC value for a subset of observers is calculated by comparing the fixation maps generated from the gaze points of those observers against the fixation map generated from the remaining observers. In other words, the process explained in sub-section 4.4 is repeated for the $i$ and $N$-$i$ observers to generate the fixation maps. The AUC values resulting from comparing these fixation maps are reported in Fig. 8. To find an estimate of the infinity-vs-infinity case, we follow a same method used in Ref. 31 by fitting a curve with the following form to this graph:

$$AUC(x) = ax^b + c \tag{2}$$

where $a$, $b$, and $c$ are constant coefficients and $x$ is the number of observers. The AUC for an infinite number of humans is extrapolated from Fig. 8. In other words, the limit of the AUC function when the number of observers approaches infinity is an approximation of the upper bound. The same procedure is repeated for other metrics to find the upper/lower bounds of each metric. The values of the parameters in (2) for different metrics along with their corresponding 95% confidence intervals (CIs) are reported in Table 3.

**Table 3.** Curve Fitting Parameters for Different Metrics

| Metric | a (CI) | b (CI) | c (CI) | One human | Limit at infinite number of humans |
|---|---|---|---|---|---|
| AUC | -0.30 (-0.32,-0.28) | -0.48 (-0.50,-0.47) | 0.9921 (0.98,1) | 0.7033 | 0.9921 |
| sAUC | -0.22 (-0.24,-0.21) | -0.54 (-0.57,-0.50) | 0.9908 (0.98,1) | 0.7379 | 0.9908 |
| NSS | -2.4 (-2.5,-2.1) | -0.49 (-0.51,-0.48) | 4.2524 (4.25,4.27) | 2.1140 | 4.2524 |
| PCC | -0.55 (-0.57,-0.51) | -0.66 (-0.70,-0.64) | 0.9968 (0.99,1) | 0.4995 | 0.9968 |
| SIM | -0.36 (-0.40,-0.33) | -0.37 (-0.39,-0.35) | 0.9511 (0.94,0.96) | 0.4651 | 0.9511 |
| KLD | 1.32 (1.21,1.44) | -0.36 (-0.42,-0.32) | 0 (0,0.01) | 0.2232 | 0 |
| EMD | 0.28 (0.25,0.31) | -0.22 (-0.24,-0.20) | 0.03 (0.02,0.04) | 0.8884 | 0.03 |

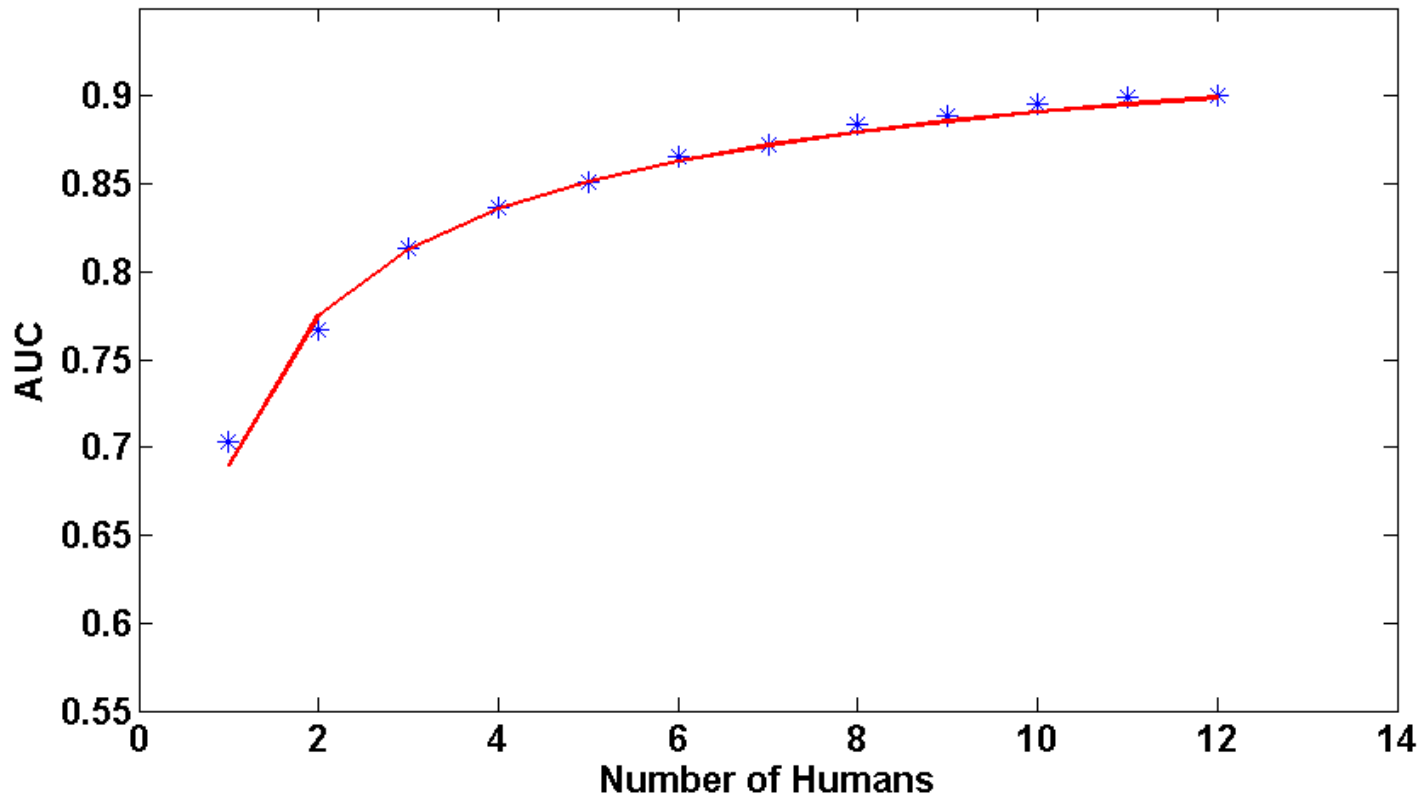

**Fig. 8.** AUC appears to converge to 0.99 for an infinite number of human observers

*5.3 Influence of the presentation order on the FDMs*

It was mentioned in sub-section 4.3 that during our eye-tracking experiments, the order of the 2D and 3D videos was switched for every subject to ensure that the subjects were not biased towards the order of the presentation material, so counter balance is achieved. This sub-section verifies our claim quantitatively. To this end, inspired by Nemoto et al.[57], we partition the FDMs into four groups: 1) 3D FDMs-First: FDMs from the 3D videos from those subjects who first watched the 3D videos and then the 2D ones, 2) 2D FDMs-First: FDMs from the 2D videos from those subjects who first watched the 2D content, 3) 2D FDMs-Second: FDMs from the 2D videos from those subjects who watched 2D videos after the 3D ones, and 4) 3D FDMs-Second: FDMs resulted from the 3D videos from those subjects who watched first the 2D videos and then the 3D ones.

First, we compute the similarity between the 3D FDMs-First vs. 3D FDMs-Second (AUC = 0.80) as well as the similarity between the 2D FDMs-First vs. 2D FDMs-Second (AUC = 0.86) fixation maps. The high similarity values show that regardless of viewing the 3D videos (or 2D) at the first or second session, there is a strong correlation between each corresponding pair of FDMs. In addition to the similarity values, we also perform a statistical testing by measuring the P-values, as shown in Table 4. The values in Table 4 show that there is a significant difference between the FDMs of 2D and 3D videos, and that the presentation order has no influence on the results.

**Table 4.** P-values when comparing different meaningful FDM partitions

| **Scenario** | **3D-1st vs. 3D-2nd** | | | **2D-1st vs. 3D-1st** | | | **2D-2nd vs. 3D-2nd** | | |
|---|---|---|---|---|---|---|---|---|---|
| | **sAUC** | **KLD** | **NSS** | **sAUC** | **KLD** | **NSS** | **sAUC** | **KLD** | **NSS** |
| **2D-1st vs. 2D-2nd** | 0 | 0.001 | 0 | 0.001 | 0.001 | 0 | 0.001 | 0.001 | 0.001 |
| **3D-1st vs. 3D-2nd** | | | | 0.001 | 0.002 | 0.001 | 0.001 | 0.002 | 0.001 |
| **2D-1st vs. 3D-1st** | | | | | | | 0.38 | 0.46 | 0.29 |

*5.4 Comparing the Performance of State-of-the-Art VAMs*

This section provides a comparison between the performance of the existing 2D and 3D visual attention models over our stereoscopic video dataset. Currently, we have considered 50 different models. Any new model can be submitted via our web page and will be added to the benchmark[55]. The selected models were originally designed for 2D/3D image/video saliency detection. The following VAMs are considered in our comparisons (in alphabetical order): AIM (Attention based on Information Maximization)[25], AIR (Saliency Detection in the Compressed Domain for Adaptive Image Retargeting)[11], Bayesian (Saliency via Low and Mid-Level Cues)[108], BMS (Boolean Map based Saliency)[106], Chamaret[79], Context-Aware saliency[109], CovSal (Visual saliency estimation by nonlinearly integrating features using region covariances)[110], DCST (Video Saliency Detection via Dynamic Consistent Spatio-Temporal Attention Modelling)[111], LBVS-3D (Learning based Visual Saliency prediction for 3D video)[82], DSM (Depth Saliency Map)[51], Fan[112], Fang[58], FT (Frequency-tuned salient region detection)[113], GBVS (Graph-Based Visual Saliency)[114], HC (Global Contrast based Salient Region Detection)[115], HDCT (Salient Region Detection via High-dimensional Color Transform)[116], HSaliency (Hierarchical Saliency Detection)[117], HVSAS (Bottom-up Saliency Detection Model Based on Human Visual Sensitivity and Amplitude Spectrum)[118], ImgManipulate (Saliency for Image Manipulation)[119], Itti (including the motion features)[75], Jiang[77], Ju[120], Judd[24], Kocberber[121], LC video (Visual attention detection in video sequences using spatiotemporal cues)[122], Ma[123], Manifold Ranking[124], Self-resemblance video[125], Niu[59], Ouerhani[72], Park[74], PCA[126], Rahtu[127], Rare (rarity-based saliency detection)[128], RC (Saliency using Region Contrast)[115], RCSS (saliency map based on sampling an image into random rectangular regions of interest)[129], FES (Fast and Efficient Saliency)[130], SDSP (saliency detection method by combining simple priors)[131], Self-resemblance (for static scene only)[125], SIM-saliency (Saliency Estimation using a non-parametric vision model)[132], Spectral Saliency[133], SR (spectral residual approach)[78], SUN (Saliency Using Natural statistics)[105], SWD (Visual Saliency Detection by Spatially Weighted Dissimilarity)[134], Torralba[135], Coria[81], Yubing[136], and Zhang[80]. Note that for the 2D image/video models, the saliency is computed only for one of the views (right view). Note that each model may have had its own objectives and history. However, for the sake of completeness, we present an evaluation of these models using our eye-tracking dataset.

Moreover, we compare the performance of the different visual attention models with those of the four baseline models mentioned in sub-section 5.2.

It is recommended (and is common practice) to perform histogram matching between a saliency map and the corresponding human fixation map[31], so that the saliency evaluation metrics attain more meaningful and fair comparisons. Note that histogram matching forces saliency maps to place the majority of the salient locations at the fixated regions.

In addition to histogram matching, blurring and adding center bias is a common practice for saliency prediction models as they slightly increase the performance of different models[31,104,108,137]. To account for the center-bias, a Gaussian disk located at the center of the coordinate axes (with the same size as the saliency map) is added to each saliency map as follows:

$$S' = w.S + (1-w).S_{center} \tag{3}$$

where *w* specifies the weight of the center-bias. To account for the blurring effect, each saliency map is convolved with a Gaussian (of the same size). The values of the weight of center-bias (*w*) are swept between 0 and 1, while the values of the standard deviation for the center-bias and blurring are swept between 0.1 and 500 (pixels). These three parameters are exhaustively swept over a wide range of values, and the set of parameters that results in the highest AUC is selected. Histogram matching, blurring, and adding a center-bias generally can increase the AUC performance by several percentage points. Among these, histogram matching has the highest impact on the AUC. Blurring and center-bias has only minor influence on the performance of different VAMs when using our dataset. One reason is that we don't have severe amounts of center-bias in our eye-tracking experiments (see Fig. 7).

Table 5 shows the performance of different 3D VAMs with respect to several saliency prediction metrics. In order to compare the accuracy using all of the metrics, we assign a rank to each model using each metric independently and then sorting the resulting ranks. In other words, assuming that various metrics are of the same importance, average of the ranks demonstrates how well each VAM performs with respect to all of the metrics. As mentioned in sub-section 5.1, most performance metrics are similar and using sAUC, KLD, and one other metric among NSS, PLCC, SROCC, and SIM provides a fair comparison. To avoid the introduction of bias in the ranking, we use only sAUC, KLD, and NSS for ranking the performance of various VAMs.

**Table 5.** Performance evaluation of different 3D VAMs using our eye-tracking dataset of stereoscopic videos

| Model | AUC | sAUC | EMD | SIM | PCC | KLD | NSS | Simulation Time (sec) | Average Rank* | Type |
|---|---|---|---|---|---|---|---|---|---|---|
| Infinite humans | 0.9921 | 0.9908 | 0.03 | 0.9511 | 0.9968 | 0 | 4.2524 | | 1 | |
| LBVS-3D [82] | 0.7243 | 0.7795 | 0.4528 | 0.2966 | 0.2620 | 0.1289 | 1.4167 | 73.39 | 2.33 | 3D video |
| LBVS-3D (static**) | 0.6833 | 0.7091 | 0.5310 | 0.2544 | 0.2376 | 0.1963 | 1.1782 | 30.55 | 3.33 | 3D image |
| One human | 0.7033 | 0.7379 | 0.8884 | 0.4651 | 0.4995 | 0.2232 | 2.1140 | | 5.33 | |
| Fang [58] | 0.6655 | 0.6915 | 0.6676 | 0.2229 | 0.1987 | 0.2165 | 1.0380 | 3.25 | 5.33 | 3D image |
| Coria [81] | 0.6584 | 0.6843 | 0.6568 | 0.2346 | 0.1417 | 0.2238 | 1.1361 | 3.03 | 8 | 3D video |
| Chamaret [79] | 0.6669 | 0.6787 | 0.7568 | 0.2089 | 0.1568 | 0.2253 | 0.9056 | 64.62 | 8.33 | 3D video |
| Park [74] | 0.6391 | 0.6346 | 0.8081 | 0.1841 | 0.1022 | 0.2198 | 0.7783 | 1.68 | 9 | 3D image |
| Ouerhani [72] | 0.6224 | 0.6456 | 0.8768 | 0.1934 | 0.0967 | 0.2179 | 0.5459 | 7.21 | 9.33 | 3D image |
| Fan [112] | 0.6349 | 0.6330 | 0.9014 | 0.1879 | 0.0856 | 0.2116 | 0.4185 | 128.98 | 9.33 | 3D image |
| Niu [59] | 0.6078 | 0.6124 | 0.9339 | 0.1726 | 0.1208 | 0.2227 | 0.3334 | 165.82 | 9.67 | 3D image |
| Ju [120] | 0.5811 | 0.5948 | 1.0330 | 0.1623 | 0.0827 | 0.2109 | 0.2778 | 2.05 | 10.67 | 3D image |
| Jiang [77] | 0.6158 | 0.6089 | 0.9949 | 0.1934 | 0.1211 | 0.2326 | 0.3656 | 1.25 | 11.33 | 3D image |
| Center | 0.5709 | 0.5999 | 0.6536 | 0.2128 | 0.1104 | 0.2445 | 0.6524 | 0.06 | 13 | |
| Zhang [80] | 0.5699 | 0.5754 | 1.0970 | 0.1528 | 0.0563 | 0.2293 | 0.2111 | 0.73 | 14.33 | 3D image |
| Chance | 0.5 | 0.5 | 1.1140 | 0.1421 | 0 | 0.2393 | 0.0789 | 0.072 | 15.67 | |

* Ranking is done using only sAUC, KLD, and NSS.
** Motion features are excluded.

Table 6 shows the performance of 2D VAMs against the FDM results of 2D and 3D eye-tracking experiments. We compare the saliency maps produced by these models with the fixation maps resulted from eye-tracking experiments on 2D and 3D video data. For each model, P-values are calculated to measure the statistical difference between its ability to distinguish between 2D and 3D FDMs.

It is worth noting that an early approach in designing 3D VAMs was to directly integrate the saliency maps produced by 2D VAMs with the generated/captured depth maps. We investigate the performance of 2D VAMs when depth maps are combined with the saliency maps.

Table 7 shows the results of this investigation. It is observed from Table 7 that using depth maps usually improves the saliency prediction performance. However, the improvements are sometimes negligible.

### *5.5 Complexity of Different VAMs*

The mathematical definition of the complexity of an algorithm involves calculating the number of operations. Due to the complex structure of most of the visual attention models, it is not possible to calculate their mathematical complexity. Instead, the only feasible solution to compare the complexity of different algorithms is to compare their simulation times. To this end, we used a workstation with i7 CPU and 18 GBs of memory to perform complexity measurements. Note that it was ensured that no other program was running during the measurement process. The results of complexity measurements are reported in Table 5 and 6.

### *5.6 Discussions*

The performance evaluations in Table 5 and Table 6 show that the model with the highest performance takes into account high-level features (person, car, and etc.). This confirms the importance of high-level saliency features in human visual attention.

**Table 6.** Performance evaluation of different 2D VAMs using our eye-tracking dataset of stereoscopic videos: The performance of each 2D VAM is calculated against 2D and 3D FDMs, respectively. P-values are rounded to two digits of floating point.

| Models \ Metrics | sAUC | | | KLD | | | NSS | | | Simulation Time (sec) | Average Rank* | Type |
|---|---|---|---|---|---|---|---|---|---|---|---|---|
| | 2D Saliency | 3D Saliency | P-value | 2D Saliency | 3D Saliency | P-value | 2D Saliency | 3D Saliency | P-value | | | |
| Infinite humans | 0.9915 | 0.9908 | | 0 | 0 | | 4.1119 | 4.2524 | | 0 | 1 | |
| Judd [24] | 0.7374 | 0.6619 | < 0.01 | 0.2037 | 0.2189 | 0.04 | 1.1424 | 0.9066 | < 0.01 | 58 | 3.3 | 2D image |
| One human | 0.8063 | 0.7379 | | 0.2014 | 0.2232 | | 2.2967 | 2.1140 | | 0 | 4.7 | |
| CovSal [110] | 0.7120 | 0.6605 | < 0.01 | 0.2144 | 0.2241 | 0.02 | 0.9823 | 0.8911 | < 0.01 | 17.9 | 6.7 | 2D image |
| DCST [111] | 0.6523 | 0.6454 | 0.04 | 0.2098 | 0.2131 | 0.01 | 0.7569 | 0.7465 | 0.02 | 146.7 | 8.7 | 2D video |
| Spectral saliency [133] | 0.6328 | 0.6316 | 0.78 | 0.2258 | 0.2219 | 0.84 | 0.8931 | 0.8612 | 0.14 | 0.34 | 9.3 | 2D image |
| Itti [75] | 0.6458 | 0.6432 | 0.93 | 0.2196 | 0.2222 | 0.39 | 0.7908 | 0.7733 | 0.18 | 4.05 | 10 | 2D video |
| SWD [134] | 0.7009 | 0.6533 | < 0.01 | 0.2225 | 0.2448 | < 0.01 | 0.9800 | 0.8894 | < 0.01 | 11.37 | 12 | 2D image |
| AIR [11] | 0.6589 | 0.6488 | 0.09 | 0.2338 | 0.2350 | 0.51 | 0.8769 | 0.8015 | 0.02 | 7.58 | 12.3 | 2D image |
| Kocberber [121] | 0.6551 | 0.6445 | 0.16 | 0.2148 | 0.2289 | 0.08 | 0.7510 | 0.7496 | 0.66 | 122 | 12.3 | 2D video |
| AIM [25] | 0.6369 | 0.6312 | 0.04 | 0.2299 | 0.2218 | 0.02 | 0.6905 | 0.6835 | 0.01 | 25.88 | 13 | 2D image |
| Rahtu [127] | 0.6354 | 0.6232 | 0.01 | 0.2167 | 0.2216 | 0.04 | 0.6804 | 0.6792 | 0.59 | 4.84 | 13.7 | 2D video |
| BMS [106] | 0.7222 | 0.6538 | < 0.01 | 0.2018 | 0.2595 | < 0.01 | 1.0052 | 0.8843 | < 0.01 | 10 | 14.3 | 2D image |
| RC [115] | 0.6376 | 0.6280 | 0.03 | 0.2232 | 0.2305 | 0.04 | 0.7378 | 0.7125 | 0.02 | 2.2 | 15 | 2D image |
| FES [130] | 0.6228 | 0.6136 | 0.02 | 0.2198 | 0.2229 | 0.02 | 0.7123 | 0.7092 | 0.01 | 0.82 | 15.3 | 2D image |
| Rare [128] | 0.6233 | 0.6149 | 0.08 | 0.2438 | 0.2499 | 0.41 | 0.9991 | 0.8913 | 0.02 | 1.52 | 17 | 2D image |
| Yubing [136] | 0.6265 | 0.6202 | 0.28 | 0.2307 | 0.2343 | 0.56 | 0.6958 | 0.6723 | 0.02 | 40.1 | 18.3 | 2D video |
| GBVS [114] | 0.6472 | 0.6323 | 0.05 | 0.2715 | 0.2903 | 0.01 | 0.8649 | 0.8111 | 0.01 | 2 | 19 | 2D image |
| ImgManipulate [119] | 0.6277 | 0.6216 | 0.75 | 0.2546 | 0.2634 | 0.04 | 0.7824 | 0.7518 | 0.01 | 12 | 20.3 | 2D image |
| LC [122] | 0.5773 | 0.5989 | 0.01 | 0.2390 | 0.2269 | < 0.001 | 0.6584 | 0.6481 | < 0.001 | 2.3 | 20.7 | 2D video |
| HDCT [116] | 0.6218 | 0.6106 | < 0.01 | 0.2366 | 0.2416 | 0.01 | 0.6699 | 0.6576 | < 0.01 | 153.24 | 21.7 | 2D image |
| Self-resemblance [125] | 0.6189 | 0.6035 | 0.01 | 0.2346 | 0.2460 | < 0.01 | 0.6942 | 0.6894 | 0.03 | 1.29 | 22 | 2D video |
| Context Aware [109] | 0.6146 | 0.5989 | < 0.01 | 0.2388 | 0.2454 | < 0.01 | 0.6843 | 0.6646 | < 0.01 | 24.16 | 23.3 | 2D image |
| Bayesian [108] | 0.5819 | 0.5761 | 0.05 | 0.2205 | 0.2249 | 0.15 | 0.5769 | 0.5937 | 0.02 | 149.24 | 23.7 | 2D image |
| PCA [126] | 0.6044 | 0.5816 | 0.03 | 0.2541 | 0.2666 | 0.04 | 0.9411 | 0.8678 | < 0.01 | 17.88 | 24 | 2D image |
| HSaliency [117] | 0.5948 | 0.5899 | 0.01 | 0.2389 | 0.2444 | 0.04 | 0.6229 | 0.6138 | 0.03 | 14.28 | 25.3 | 2D image |
| Center | 0.6043 | 0.5898 | < 0.01 | 0.2675 | 0.2848 | < 0.01 | 0.7452 | 0.7120 | < 0.01 | 0.06 | 26.3 | |
| SR [78] | 0.5790 | 0.5673 | 0.01 | 0.2250 | 0.2359 | < 0.01 | 0.4780 | 0.4851 | 0.04 | 1.9 | 26.7 | 2D image |
| Manifold Ranking [124] | 0.6092 | 0.5935 | 0.01 | 0.3211 | 0.3685 | < 0.01 | 0.6632 | 0.6538 | 0.01 | 12.94 | 28.3 | 2D image |

**Table 6.** Performance evaluation of different 2D VAMs using our eye-tracking dataset of stereoscopic videos

| Models \ Metrics | sAUC | | | KLD | | | NSS | | | Simulation Time (sec) | Average Rank* | Type |
|---|---|---|---|---|---|---|---|---|---|---|---|---|
| | 2D Saliency | 3D Saliency | P-value | 2D Saliency | 3D Saliency | P-value | 2D Saliency | 3D Saliency | P-value | | | |
| HVSAS [118] | 0.5638 | 0.5545 | 0.04 | 0.2309 | 0.2343 | 0.04 | .2033 | 0.1932 | 0.11 | 24.6 | 28.7 | 2D image |
| FT [113] | 0.5699 | 0.5664 | 0.91 | 0.2411 | 0.2434 | 0.77 | 0.3921 | 0.3846 | 0.15 | 2 | 29.7 | 2D image |
| Torralba [135] | 0.5856 | 0.5742 | 0.01 | 0.2538 | 0.2496 | 0.01 | 0.5027 | 0.5358 | 0.01 | 3.38 | 30 | 2D image |
| Ma [123] | 0.5898 | 0.5834 | 0.42 | 0.2450 | 0.2532 | 0.04 | 0.4993 | 0.4954 | 0.40 | 39.12 | 30 | 2D video |
| SIM [132] | 0.5565 | 0.5442 | 0.10 | 0.2378 | 0.2344 | 0.48 | 0.2657 | 0.2545 | 0.45 | 12.2 | 30 | 2D image |
| RCSS [129] | 0.5942 | 0.5926 | 0.94 | 1.1034 | 1.0048 | 0.47 | 0.5531 | 0.5435 | 0.18 | 15.86 | 30.7 | 2D image |
| SDSP [131] | 0.5899 | 0.5857 | 0.55 | 0.9485 | 0.9935 | 0.63 | 0.4565 | 0.4353 | 0.04 | 0.19 | 31 | 2D image |
| HC [115] | 0.5632 | 0.5477 | < 0.01 | 0.2435 | 0.2463 | 0.25 | 0.2757 | 0.2664 | 0.01 | 2.1 | 33 | 2D image |
| SUN [105] | 0.5658 | 0.5537 | 0.01 | 0.2608 | 0.2647 | 0.04 | 0.1218 | 0.1169 | 0.05 | 3.07 | 33 | 2D image |
| Chance | 0.5 | 0.5 | 1 | 0.3348 | 0.3532 | 0.96 | 0.0011 | 0.0020 | 1 | 0.072 | 35.7 | |

* Ranking is done using only sAUC, KLD, and NSS, on the 3D saliency prediction performance of different 2D VAMs.

Moreover, 3D visual attention models in general do not show promising performance over our dataset (the highest AUC resulting from the existing VAMs is around 0.72). Comparing the performance of 2D VAMs in saliency prediction for 2D and 3D videos (see Table 6) shows that the metric values are significantly lower for the case of 3D. This was expected to some extent, since saliency prediction in 3D is much more difficult due to the impact of depth perception. Our results in Table 7 show that even tough using depth information along with 2D VAMs can potentially improve the saliency prediction accuracy, but 2D VAMs plus depth prior do not outperform the 3D VAMs. In addition, Table 5 and Table 6 show that predictions made by one human observer are not representative of an infinite number of observers, which means there are more complex visual attributes involved in 3D perception. In conclusion, the above results clearly show that there is still a need for designing more accurate 3D visual attention models.

Table 7. Performance evaluation of different 2D VAMs for saliency prediction on stereoscopic videos when they are integrated with depth maps: The values in this table show the improvements (or accuracy reduction if there is no improvement) achieved when depth maps are used.*

| Models \ Metrics | sAUC | | KLD | | NSS | |
|---|---|---|---|---|---|---|
| | multiplication | linear combination | multiplication | linear combination | multiplication | linear combination |
| Judd [24] | -0.0227 | +0.0137 | +0.0217 | -0.0094 | -0.1263 | +0.0743 |
| CovSal [110] | -0.0301 | +0.0146 | +0.0286 | -0.0096 | -0.1151 | +0.0756 |
| DCST [111] | -0.0273 | +0.0163 | +0.0135 | -0.0105 | -0.0947 | +0.0833 |
| Spectral saliency [133] | -0.0199 | +0.0274 | +0.0253 | -0.0176 | -0.0673 | +0.1405 |
| Itti [75] | +0.0082 | +0.0193 | +0.0029 | -0.0143 | +0.0652 | +0.1012 |
| SWD [134] | -0.0099 | +0.0194 | +0.0201 | -0.0151 | -0.1042 | +0.1043 |
| AIR [11] | -0.0101 | +0.0221 | +0.0193 | -0.0166 | -0.0439 | +0.1119 |
| Kocberber [121] | +0.0010 | +0.0243 | -0.0004 | -0.0163 | +0.0065 | +0.1240 |
| AIM [25] | +0.0012 | +0.0165 | -0.0099 | -0.0137 | +0.0130 | +0.0847 |
| Rahtu [127] | +0.0018 | +0.0258 | -0.0052 | -0.0158 | +0.0352 | +0.1222 |
| BMS [106] | +0.0019 | +0.0104 | -0.0009 | -0.0076 | +0.0471 | +0.0054 |
| RC [115] | +0.0022 | +0.0226 | -0.0037 | -0.0159 | +0.0518 | +0.1179 |
| FES [130] | +0.0031 | +0.0221 | -0.0009 | -0.0157 | +0.0394 | +0.1107 |
| Rare [128] | +0.0038 | +0.0194 | -0.0021 | -0.0148 | +0.0289 | +0.1005 |
| Yubing [136] | +0.0045 | +0.0336 | -0.0033 | -0.0212 | +0.0649 | +0.1668 |
| GBVS [114] | +0.0073 | +0.0241 | -0.0019 | -0.0166 | +0.0533 | +0.1243 |
| ImgManipulate [119] | +0.0015 | +0.0243 | -0.0005 | -0.0163 | +0.0115 | +0.1196 |
| LC [122] | +0.0122 | +0.0374 | -0.0201 | -0.0225 | +0.1343 | +0.1924 |
| HDCT [116] | +0.0075 | +0.0323 | -0.0005 | -0.0296 | +0.0972 | +0.1693 |
| Self-resemblance [125] | +0.0048 | +0.0493 | -0.0003 | -0.0312 | +0.0447 | +0.2045 |
| Context Aware [109] | +0.0023 | +0.0369 | -0.0001 | -0.0274 | +0.0228 | +0.1768 |
| Bayesian [108] | +0.0010 | +0.0325 | +0.0008 | -0.0178 | +0.0084 | +0.1665 |
| PCA [126] | +0.0080 | +0.0291 | -0.0093 | -0.0311 | +0.0768 | +0.1535 |
| HSaliency [117] | +0.0020 | +0.0241 | -0.0009 | -0.0188 | +0.0249 | +0.1207 |
| SR [78] | +0.0173 | +0.0434 | -0.0193 | -0.0244 | +0.1783 | +0.2132 |
| Manifold Ranking [124] | +0.0090 | +0.0258 | -0.0553 | -0.0160 | +0.0932 | +0.1263 |
| HVSAS [118] | +0.0136 | +0.0559 | -0.0100 | -0.0228 | +0.1554 | +0.2385 |
| FT [113] | +0.0111 | +0.0453 | -0.0093 | -0.0273 | +0.1426 | +0.2211 |
| Torralba [135] | +0.0093 | +0.0266 | -0.0042 | -0.0296 | +0.1077 | +0.1490 |
| Ma [123] | +0.0099 | +0.0213 | -0.0036 | -0.0199 | +0.0979 | +0.1023 |
| SIM [132] | +0.0154 | +0.0555 | -0.0126 | -0.0264 | +0.1723 | +0.2776 |
| RCSS [129] | +0.0115 | +0.0258 | -0.0164 | -0.0126 | +0.1545 | +0.1266 |
| SDSP [131] | +0.0094 | +0.0198 | -0.0735 | -0.0135 | +0.1019 | +0.1034 |
| HC [115] | +0.0200 | +0.0564 | -0.0189 | -0.0333 | +0.1910 | +0.2765 |
| SUN [105] | +0.0224 | +0.0279 | -0.0204 | -0.0307 | +0.2138 | +0.1445 |

* The values of the weights for the linear combination are selected such that for each model the AUC is maximized between that model and the FMDs of the training dataset. Moreover, when using depth maps as weights, they are first normalized to 0-1.

## *5.7 Performance of Different VAMs When Applied to Videos with no Outstanding Objects*

This sub-section investigates how having outstanding objects such as humans or cars in a scene can change the performance of 3D VAMs. To this end, only a subset of the 3D video dataset containing 22 videos are used. These videos do not include any high level salient object such as humans, vehicles, text, animals, or etc. Table 8 shows the performance evaluation results when only the mentioned subset of videos is used. The results in Table 8 show the ability of different VAMs in bottom-up saliency prediction, where no high level attribute is present in the scenes. It is observed from Table 8 that using different test videos in general changes the performance the VAMs, however, the changes are usually not significant. Some models performed better in the presence of high level cues, which might be due to the fact that they use those cues for saliency prediction. Some models perform worse when high level salient objects are present. This might be because these models are only using low level cues, or they cannot efficiently locate the high level salient objects (e.g. having trouble identifying humans or cars if they are of small size).

## 6 Conclusions

The lack of benchmark eye-tracking datasets for stereoscopic videos has slowed down the progress in developing efficient 3D visual attention models. This paper introduces a benchmark stereoscopic video dataset coupled with the data from a free-viewing eye-tracking experiment, as a step towards facilitating the design of advanced 3D visual attention models. Our dataset contains the eye fixation points collected from 61 stereoscopic videos. We evaluate the

**Table 8.** Performance evaluation of different 3D VAMs when videos with only low level saliency cues are used

| Model | sAUC | KLD | NSS |
|---|---|---|---|
| Infinite humans | 0.9924 | 0 | 4.1983 |
| LBVS-3D [82] | 0.7508 | 0.1414 | 1.3782 |
| One human | 0.7388 | 0.2119 | 2.2248 |
| LBVS-3D (static*) | 0.6992 | 0.1998 | 1.1457 |
| Fang [58] | 0.6973 | 0.2228 | 1.0982 |
| Coria [81] | 0.6665 | 0.2438 | 1.0088 |
| Chamaret [79] | 0.6671 | 0.2308 | 0.9257 |
| Niu [59] | 0.6587 | 0.2178 | 0.5974 |
| Park [74] | 0.6449 | 0.2177 | 0.7628 |
| Ouerhani [72] | 0.6411 | 0.2265 | 0.5319 |
| Fan [112] | 0.6166 | 0.2294 | 0.4093 |
| Ju [120] | 0.5933 | 0.2301 | 0.3266 |
| Center | 0.5896 | 0.2477 | 0.5838 |
| Jiang [77] | 0.5843 | 0.2370 | 0.3309 |
| Zhang [80] | 0.5804 | 0.2299 | 0.2665 |
| Chance | 0.5 | 0.2300 | 0.0778 |

* Motion features are excluded.

performance of existing 3D saliency prediction models over the proposed dataset. Moreover, we evaluate the performance of the existing 2D VAMs against 2D and 3D eye-tracking FDMs and show that the two cases are statistically performing differently for most VAMs. A web page is designed to present the performance of existing 2D and 3D visual attention models and to submit new 3D VAMs to be compared to the currently implemented ones.

**Acknowledgements**

This work was partly supported by Natural Sciences and Engineering Research Council of Canada (NSERC) under Grant STPGP 447339-13 and the Institute for Computing Information and Cognitive Systems (ICICS) at UBC.

**Biographies**

*A. B. Dehkordi* received B.A.Sc and M.A.Sc. degrees in Electrical and Computer Engineering from the University of Tehran, Iran, in 2008 and 2011, and PhD degree in the Digital Media Lab (DML) at the University of British Columbia (UBC), Canada, in 2015. His research interests include 3D visual attention modeling, 3D TV, 3D video quality assessment, high dynamic range video, and in general various topics in digital video processing and compression.

*M. T. Pourazad* received the Ph.D. degree in Electrical and Computer Engineering from the University of British Columbia (UBC) in 2010. She is a Research Scientist in TELUS Communications Inc., and Associate Research Director at UBC (ICICS). Her field of research includes 3D video processing, 3D quality of experience and multiview video compression. Dr. Pourazad has been an active member of IEEE, the Standards Council of Canada (SCC) and the Moving Pictures Experts Group (MPEG).

*P. Nasiopoulos* is a Professor with the UBC department of Electrical and Computer Engineering, the Inaugural Holder of the Dolby Professorship in Digital Multimedia, and the current Director of the Master of Software Systems Program at UBC. Before joining UBC, he was the President of Daikin Comtec US (founder of DVD) and Executive Vice President of Sonic Solutions. Dr. Nasiopoulos is a registered Professional Engineer in British Columbia, Fellow of the Canadian Academy of Engineering, and has been an active member of the Standards Council of Canada, MPEG, ACM and IEEE.